\documentclass[english,aps,preprint,showpacs]{revtex4}
\usepackage{ae,aecompl}
\usepackage[T1]{fontenc}
\usepackage[latin9]{inputenc}
\usepackage{verbatim}
\usepackage{textcomp}
\usepackage{amsmath}
\usepackage{graphicx}
\usepackage{amssymb}
\usepackage{subfigure}
\usepackage{multirow}
\makeatletter
\usepackage{textcomp}
\usepackage{babel}
\usepackage{amsfonts}

\begin{document}

\title{Robust dynamical decoupling}
\author{Alexandre M. Souza}
\email{alexandre@e3.physik.uni-dortmund.de}
\affiliation{Fakult\"at Physik, Technische Universit\"at Dortmund, D-44221 Dortmund,
Germany.}

\author{Gonzalo A. \'Alvarez}
\email{gonzalo.alvarez@tu-dortmund.de}
\affiliation{Fakult\"at Physik, Technische Universit\"at Dortmund, D-44221 Dortmund,
Germany.}

\author{Dieter Suter}
\email{Dieter.Suter@tu-dortmund.de}
\affiliation{Fakult\"at Physik, Technische Universit\"at Dortmund, D-44221 Dortmund,
Germany.}



\begin{abstract}
 Quantum computers, which process information encoded in
quantum mechanical systems, hold the potential to solve some of the  hardest computational problems.
A substantial
obstacle for the further
development of quantum computers is the fact that the life time of
quantum information is usually
too short to allow practical computation. A promising method to
increase the life time, known as
dynamical decoupling, consists of applying a periodic series of
inversion pulses to the quantum bits.
In the present review, we give an overview of this technique and compare
different pulse sequences proposed earlier.
We show that pulse imperfections, which are always present in experimental implementations,
limit the performance of dynamical decoupling.
The loss of coherence due to the accumulation of pulse errors can 
even exceed the perturbation from the environment.
This effect can be largely eliminated by a judicious design of pulses and sequences.
The corresponding sequences are largely immune to pulse imperfections
and provide an increase of the coherence time of the system by several orders of magnitude.
\end{abstract}

\maketitle
\section{Introduction}

During the last decade, it was shown that
quantum mechanical systems have  the potential for processing information more efficiently
than classical systems \cite{Shor1994,DiVincenzo1995,Nielsen00,BD2000}.
However, it remains difficult to realize this potential because
quantum systems are extremely sensitive to perturbations. These perturbations
come from external degrees of freedom or from the finite precision with 
which the systems can be realized and controlled by external fields.
This loss of  quantum information to the environment is called decoherence \cite{Zurek03}. 
Different results  show that a state is more sensitive
to decoherence as the number of qubits increases \cite{Suter04,Suter06,Krojanski2006,Cory06,Lovric2007,sanchez_time_2007,Doronin2011,Zobov2011}.
This is also manifested by the impossibility to time reverse a quantum
evolution when an initially localized excitation spreads over a system
\cite{Pines70,Zhang1992,Usaj98,PhysicaA}. As the information is distributed
over a increasing number of qubits, the evolution becomes more sensitive to the perturbation
\cite{PhysicaA,JalPas01}. In a similar vein, this sensitivity with
the systems size limits the distance over which one can transfer information
or analogously limits the number of qubits that one 
can control reliably \cite{alvarez_nmr_2010,Alvarez2011}. This is manifested as
a localization effect for the quantum information \cite{Anderson1958,Pomeransky2004,Chiara2005,Burrell2007,Keating2007,Apollaro2007,Allcock2009,alvarez_nmr_2010,Alvarez2011}.
In order to overcome these limitations for allowing quantum information
processing with large number of qubits, methods for reducing
the decoherence effects have to be developed.

One can tackle the problem by correcting the errors generated by the
perturbations, but this is only possible if the perturbation is small
enough to keep the quantumness of the system \cite{error1,3921,6581}. Therefore,
one needs first to reduce the perturbation effects. 
The pioneering strategies for reducing decoherence were introduced in
the Nuclear Magnetic Resonance (NMR) community, in particular by Erwin Hahn
who showed that  inverting
a spin-1/2 system (a qubit) corresponds to an effective change of the sign
of the perturbation Hamiltonian and therefore generates a time reversal of the corresponding
evolution \cite{Hahn50b}. 
This leads to the formation of an echo
that later  was formalized as a Loschmidt echo \cite{JalPas01}. 
These manipulations were extended to the so-called decoupling methods \cite{Carr1954,Meiboom1958,Waugh1968,Maudsley1986,Gullion1990},
which disconnect effectively the environment.

In the context of this review, we describe the environment as a spin-bath, without loss of generality. 
Considering spins 1/2 as qubits, two different types of decoupling
methods can be distinguished. 
In the first case, the qubit  system is well distinguished from the environment.
Its energy level splitting differs significantly from that of the bath.
As a result, the
coupling between them is much smaller than the difference of their energy level splittings.
The interaction can then always be approximated by an Ising-type ($zz$) interaction,
which causes dephasing of the system qubit but no qubit flips.
The decoupling methods required for these cases are called heteronuclear decoupling
 within the NMR community and they can involve manipulation
only on the spin-system \cite{Hahn50b,Carr1954,Meiboom1958,Maudsley1986,Gullion1990}
or only at the environment \cite{waugh69}.

In the second case, the system and the environment have similar energy level splittings.
This is the case of a homonuclear system where the general form of the coupling
must be retained and it can induce flips of the system qubit as well as dephasing.
In this case, decoupling will generally affect the 
complete system plus ``bath'' \cite{waugh68b,mansfield73,rhim73,rhim74,burum79,Burum81}.

In this review we focus on decoupling the system from the environment by applying
control pulses only to the system. 
During the last years this technique
has gathered a lot of interest because it requires relatively modest
resources: it requires no overhead of information encoding,
measurements or feedback. 
The method is known as dynamical decoupling (DD).
Since its initial introduction
\cite{viola_dynamical_1999},
a lot of effort has been invested to find sequences with 
improved error suppression \cite{Viola2003,khodjasteh_fault-tolerant_2005,Uhrig2007,Gordon2008,biercuk_optimized_2009,Uhrig2009,Yang2010,Clausen2010,west_near-optimal_2010,Souza2011}.
The optimal design of DD sequences depends of the different sources
of errors that have to be eliminated. The first to be considered is
the nature of the SE interaction, i.e., a pure dephasing when only 
the  $1/2$-spin operator $S_z$ is present, a pure spin-flip interaction 
($S_x$ and/or $S_y$ are present) or a general interaction with $S_x$, $S_y$ and $S_z$.
Sequences like Carr-Purcell (CP) and Carr-Purcell-Meiboom-Gill (CPMG), which use rotations around a single axis,
are useful  when only two operators of $S_x$, $S_y$ and $S_z$ are present, but in order to fight against
a general interaction, pulses along different
spatial direction have to be applied. The shortest  sequence that fulfills this 
condition is the XY-4 sequence \cite{Maudsley1986,viola_dynamical_1999}. 
An actual implementation must take into account, in addition to the above issues,
the effect of pulse imperfections \cite{Viola2003,Alvarez2010,Ryan2010,Souza2011,Wang2011,xiao2011}.
Fighting the effect of pulse errors  was in fact the original motivation for the development of the XY-4 sequence
\cite{Maudsley1986}.
Another experimental consideration is the amount of power deposited in the system,
which often must be kept small to avoid heating effect or damage to the sample.

DD technique is becoming an important tool for quantum information processing
\cite{Ardavan06,Morton2008,biercuk_optimized_2009,Du2009,Alvarez2010,Lange2010,Ryan2010,Barthel2010,Bluhm2010,Ajoy2011,Naydenov2011,Souza2011,Souza2011a}
as well as in spectroscopy \cite{Jenista2009,noise1,noise2,Alvarez2011a} and imaging \cite{magnetometer1,meriles,magnetometer2,magnetometer3}.
In most cases, the goal is to preserve a given input state, but it may also be combined
with gate operations \cite{ddgate1,ddgate2,ddgate3,ddgate4,dqc1dd}.
In many cases, experimental results  show that one of the main
limitations for improving the decoupling efficiency are the non-ideal properties of
the decoupling pulses  \cite{Alvarez2010,Ryan2010,Souza2011,Souza2011a}. Decoherence
effects due to the environment can in principle be reduced by shortening the cycle time $\tau_{c}$.
However, in reality the pulse lengths are finite and the DD cycle
time is  limited by the available hardware.
It is therefore important to find the
best finite time sequences for 
decoupling \cite{Viola2003,khodjasteh_performance_2007,Hodgson2010,UhrigLidar2010,Alvarez2010,Souza2011,Souza2011a,Ng2011}.
Furthermore, the pulses do not only have finite lengths, they also do not implement
perfect rotations.
Thus, increasing the number of pulses can result in a large overall error
that destroys the qubit coherence instead of preserving it.
As a result, the performance of the decoupling sequence may have an optimum
at a finite cycle time \cite{Alvarez2010,Souza2011,Souza2011a}.
In this review, we summarize the different DD strategies for  fighting the effect of
pulse imperfections and we show that they must be
considered in the designing of useful DD sequences. 

The paper is structured as follows: In section II we give some basics of dynamical decoupling, in section III we introduce the effects of pulse imperfections 
and in section IV we describe the strategies to fight against the imperfections effects. In the last section we draw some conclusions and perspectives.

\section{Basics of dynamical decoupling}

\subsection{The system}

We consider a single qubit $\hat{S}$ as the system that is coupled
to the bath. In a resonantly rotating frame of reference \cite{Abragam},
the free evolution Hamiltonian is 
\begin{equation}
 \widehat{\mathcal{H}}_{f}=\widehat{\mathcal{H}}_{SE}+\widehat{\mathcal{H}}_{E},
\end{equation}
where $\widehat{\mathcal{H}}_{E}$ is the environment Hamiltonian and 
\begin{equation}
 \widehat{\mathcal{H}}_{SE}=\sum_\beta \left(b_{z}^\beta \hat{E^\beta_{z}}\hat{S}_{z}+b_{y}^\beta \hat{E^\beta_{y}}\hat{S}_{y}+b_{x}^\beta \hat{E^\beta_{x}}\hat{S}_{x}\right)
\end{equation}
is a general interaction between the system and the environment. $\hat{E}^\beta_{u}$
are operators of the environment and $b^\beta_{u}$ the SE coupling
strength. 
The index $\beta$ runs over all modes of the environment.
Dephasing is due to an interaction that affects the $z$
component of the spin-system operator, and spin-flips are done trough the
$x$ and/or $y$ operators. A heteronuclear spin-spin
interaction involves a pure dephasing interaction. This type of interaction that is naturally
encountered when the system can be distinguished from the environment
can be found in a wide range of solid-state spin systems, as for example
nuclear spin systems in NMR \cite{Carr1954,Meiboom1958,Alvarez2010,Ajoy2011},
electron spins in diamonds \cite{Ryan2010}, electron spins
in quantum dots \cite{Hanson2007}, donors in silicon \cite{Kane98},
etc. In other cases when the system and environment have similar
energies, the SE interaction can include terms along the $x$, $y$ and $z$ axis.

\subsection{DD sequences with a single rotation axis}

\begin{figure}[htbp]
\vspace*{13pt}
\begin{center}
{\includegraphics[width=8.5cm]{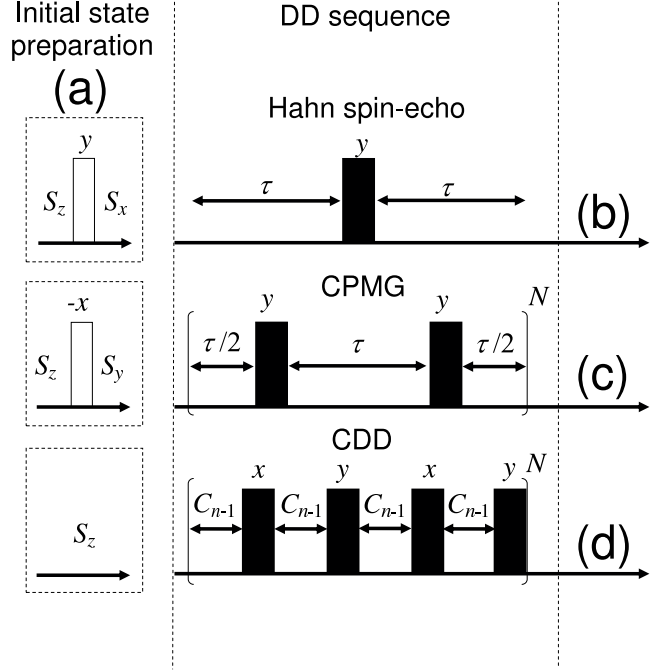}}
\end{center}
\vspace*{13pt}
\caption{\label{ddseqs}  Dynamical decoupling pulse sequences. The
empty and solid rectangles represent $90^\circ$ and $180^\circ$ pulses, respectively and
$N$ represents the number of iterations of the cycle. (a) Initial state
preparation. (b) Hahn spin-echo sequence. (c) CPMG sequence. (d) CDD sequence of order $n$, CDD$_n$ =
C$_n$ and C$_0=\tau$.}
\end{figure}

DD is achieved by iteratively applying to the system a series of stroboscopic
control pulses in cycles of period $\tau_{c}$ \cite{viola_dynamical_1999}.
Over that period, the time-averaged SE interaction can be described
by an averaged or effective Hamiltonian  \cite{Haeberlen1976}. 
The goal of DD is the elimination of
the effective SE interaction.
This can be  seen by looking at Hahn's pioneering  spin-echo
experiment \cite{Hahn50b} (See Fig. \ref{ddseqs}b). It is based on the application of a $\pi$-pulse
to the spin system at a time $\tau$ after the spins were left to evolve in the magnetic field. 
This pulse effectively changes the sign of the system-environment
(SE) interaction, in this case the Zeeman interaction with the magnetic
field. Letting the system to evolve for a refocusing period or time
reversed evolution during the same duration $\tau$ generates the
echo. 
If the magnetic field is static, the dynamics is completely reversed and the initial
state of the spin recovered. 
This is observed as the echo. 
However if the magnetic
field  fluctuates, its effect cannot be  reversed completely.
Thus, the echo amplitude decays as a function of the refocusing time
\cite{Hahn50b,Carr1954}. 
This decay contains information about the time-dependence of the environment.

To reduce the decay rate of the echo due to a time-dependent environment,
Carr and Purcell introduced a variant of the Hahn-echo sequence, where
the single $\pi$-pulse is replaced by a series of pulses separated
by intervals of duration $\tau$ \cite{Carr1954}. This CP sequence reduces the
changes induced by the environment if the pulse intervals are shorter
than the correlation time of the environment. 
However, as the number of pulses increases, pulse errors tend to accumulate.
Their combined effect can destroy the state of the system, rather than 
preserving it against the effect of the environment.
This was noticed by Meiboom and Gill who proposed a modification of the
CP sequence for compensating pulse errors, the CPMG sequence \cite{Meiboom1958}.

CP and CPMG sequences are useful only when two of the spin operators $S_x$, $S_y$ and $S_z$ are 
affected by the environment. 
They can be written as
\begin{equation}
 f_{\tau/2}\hat{Y}f_{\tau}\hat{Y}f_{\tau/2},
\end{equation}
where $f_{\tau}$ is a free evolution operator and $\hat{Y}$ is a
$\pi$ pulse around the $Y$ axis (and the analogous for $\hat{X}$). 
The difference between the CP and CPMG sequence is the orientation of the rotation axis
with respect to the initial condition.
For applications in quantum information processing, this distinction is not relevant,
since all operations have to be independent of the initial condition.
CP or CPMG is the shortest
sequence of pulses for decoupling a SE interaction that affects only two of the spin 
components $S_z$, $S_y$ and $S_z$ \cite{viola_dynamical_1999}. 

Usually the average Hamiltonian generated by DD sequences can be described
by a series expansion, such as the Magnus expansion\cite{Magnus1954}.
All the higher-order terms in this expansion describe imperfections,
which reduce the fidelity of the sequence and should be eliminated.
Improving the DD performance is therefore related to reducing the contribution
of higher order terms. 
This is closely related to efforts for developing better decoupling sequences
for NMR \cite{Waugh1968}. 
For quantum information processing (QIP), this lead to the design of sequences that make
DD more effective, such as  concatenated dynamical decoupling
\cite{khodjasteh_fault-tolerant_2005,khodjasteh_performance_2007}.
An important innovation was due to G. Uhrig  \cite{Uhrig2007}, who proposed a 
sequence with non-equidistant pulse spacings, while all the standard sequences like CPMG
are based on equidistant pulses.
The UDD sequence is defined by 
\begin{equation}
 \mathrm{UDD}_{N}=f_{\tau_{N+1}}\hat{Y}f_{\tau_{N}}\hat{Y}...\hat{Y}f_{\tau_{2}}\hat{Y}f_{\tau_{1}},
\end{equation}
where the delays $\tau_{i}=t_{i}-t_{i-1}$ are determined by the positions
\begin{equation}
t_{i}=\tau_{c}\sin^{2}\left[\frac{\pi i}{2\left(N+1\right)}\right]
\end{equation}
of the pulses
with $t_{N+1}=\tau_{c}$ and $t_{0}=0$. 
The lowest nontrivial order is equal to the CPMG sequence, UDD$_{2} = $ CPMG.

The effect of the non equidistant pulses can be discussed in the context 
of filter theory:  DD  can be considered as an environmental
noise filter, where the distribution of pulses generates different
filter shapes as a function of the frequencies.
The overlap of this filter function with the  spectral distribution of the environmental noise 
determines the decoherence rate \cite{Cywinski2008}.
Analogously, the filter shapes can be connected to diffraction patterns
induced by interferences in the time domain \cite{Ajoy2011}. The
Uhrig DD sequence was shown theoretically to be the optimal sequence for reducing low frequency
noise \cite{Uhrig2007,Cywinski2008,Uhrig2008}.
This prediction was confirmed experimentally
 \cite{biercuk_optimized_2009,Du2009,Jenista2009}.
 However, it appears that non-equidistant sequences perform better only for particular
noise spectral densities that increase for higher frequencies and have a strong
cut off.
In usual case, where the spectral density decrease smoothly
with the frequency, as usually happens with NMR spin baths, equidistant
sequences were predicted \cite{Cywinski2008,Uhrig2008,pasini_optimized_2010}
and demonstrated \cite{biercuk_optimized_2009,Alvarez2010,Lange2010,Ryan2010,Barthel2010,Ajoy2011}to
be the best option \cite{Ajoy2011}. 

This filter function description
can be traced back to previous NMR approaches \cite{Garroway1977}
and to work on universal dynamical control \cite{Kofman2001}.
Choosing the times for the pulses
leads to a variety of sequences that can be optimized according to
the spectral density of the bath \cite{Kofman2001,Gordon2008,biercuk_optimized_2009,Uys2009,Clausen2010,pasini_optimized_2010,Pan2010}.

If the SE interaction is a pure dephasing one, it is sufficient 
to apply pulses in one direction. 
However because every experimental setup
has finite precision, pulse errors create an
effective general SE 
interaction \cite{khodjasteh_performance_2007,Alvarez2010,Souza2011a}. 
In this case, it 
was shown that sequences that
are designed for general SE interaction perform better than 
1D sequences \cite{Alvarez2010,Ryan2010,Souza2011}. 

\subsection{DD sequences with multiple rotation axis }

If the system-environment interaction includes all three components of the system spin operator,
decoupling can only be achieved if the sequence includes rotations
around at least two different axes.
The first decoupling sequence of this type is the XY-4 sequence,
which alternates rotations around the x- and y-axes
(see Fig. \ref{ddseqs} d, $n=1$).
This sequence was initially
used  to eliminate the effect of pulse errors
in the CP and CPMG sequences \cite{Maudsley1986}. 
It is also the shortest sequence
for DD for general SE interactions \cite{viola_dynamical_1999}.
In quantum information processing, where we consider the initial state to be unknown,
the CP and CPMG sequences, which correspond to a train of identical $\pi$-pulses, 
are identical \cite{Alvarez2010}. 
However, their effect on the quantum state depends on the (generally unknown) initial condition:
If the initial condition is oriented along the rotation axis of the pulses,
flip angle errors of the first pulse are refocused by the second pulse. 
However, for components perpendicular to the rotation axis, 
the pulse errors of all pulses add and cause rapid decay of the coherence,
even in the absence of system-environment interactions \cite{Maudsley1986,Alvarez2010}. 
This motivated the development of the XY-4 sequence. 
In addition, pulse imperfections convert an ising-type system environment interaction 
into an effective general SE interaction \cite{khodjasteh_performance_2007,Alvarez2010,Souza2011a},
which is not eliminated by the CP/CPMG sequence, but is partially eliminated by the XY-4 sequence.
In the QIP community, the XY-4 sequence is usually referred to as 
periodic dynamical decoupling (PDD). 

The XY-4 sequence is also the building block for
concatenated DD (CDD) sequences that improve the decoupling efficiency
\cite{khodjasteh_fault-tolerant_2005,khodjasteh_performance_2007}.
The CDD scheme recursively concatenates lower order sequences to increase
the decoupling power. The CDD evolution operator for its original  version for 
a recursion order of $n$ is given by 
\begin{equation}
 \mathrm{CDD}_{n}=C_{n}=\hat{Y}C_{n-1}\hat{X}C_{n-1}\hat{Y}C_{n-1}\hat{X}C_{n-1},
\label{cdd}
\end{equation}
where $C_{0}=f_{\tau}$ and $\mathrm{CDD}_{1}$=XY-4. Figure \ref{ddseqs}
shows a general scheme for this process. Each level of concatenation
reduces the norm of the first non-vanishing order term of the Magnus
expansion of the previous level, provided that the norm was small
enough to begin with \cite{khodjasteh_fault-tolerant_2005,khodjasteh_performance_2007}.
This reduction comes at the expense of an extension of the cycle
time by a factor of four. 
If the delays between the pulses are allowed to be non-equidistant like in UDD,
it becomes possible to create hybrid sequences, such as  CUDD \cite{Uhrig2009}
and QDD \cite{west_near-optimal_2010,qdd1,qdd2}.

\section{Effects of imperfections}

Since the precision of any real operation is finite, the control fields used for decoupling introduce 
errors.
Depending on the sequence, these errors can accumulate.
If the number of pulses is large and the sequence is not properly designed,
the accumulated pulse errors can lead to severe loss of coherence than the 
effect of the environment.
Designing effective decoupling sequences that suppress environmental effects
without degrading the system, even if the control fields have errors,
requires a careful analysis of the relevant errors and appropriate strategies
for combining rotations in such a way that the errors cancel rather than accumulate.

One example of a non-ideal control pulse is its finite duration, which implies a minimum
achievable cycle time. 
The effects introduced by  finite pulse lengths have
been considered in different theoretical works 
\cite{Viola2003,khodjasteh_performance_2007,Hodgson2010}. These 
works predict that high order CDD or UDD sequences in 
general lose their advantages when the delays between pulses 
or pulse length are strongly constrained. While the limitation on the  
cycle reduces the maximal achievable DD performance, pulse errors 
can be even more destructive. 
In most cases, the 
dominant cause of errors is a deviation between the ideal 
and the actual amplitude the of control fields. The result 
of this amplitude error is that the rotation angle 
experimentally implemented deviates from $\pi$, typically by a few 
percent. Another important error occurs when the control field is not applied 
in resonance with the
transition of the qubit. This off-resonant effect 
produces a rotation in which the flip angle and the 
rotation axis deviate from the ideal ones.

\begin{figure}[htbp]
\vspace*{13pt}
\begin{center}
\begin{tabular}{ccc}
\multirow{2}{*}{{\includegraphics[width=6.5cm]{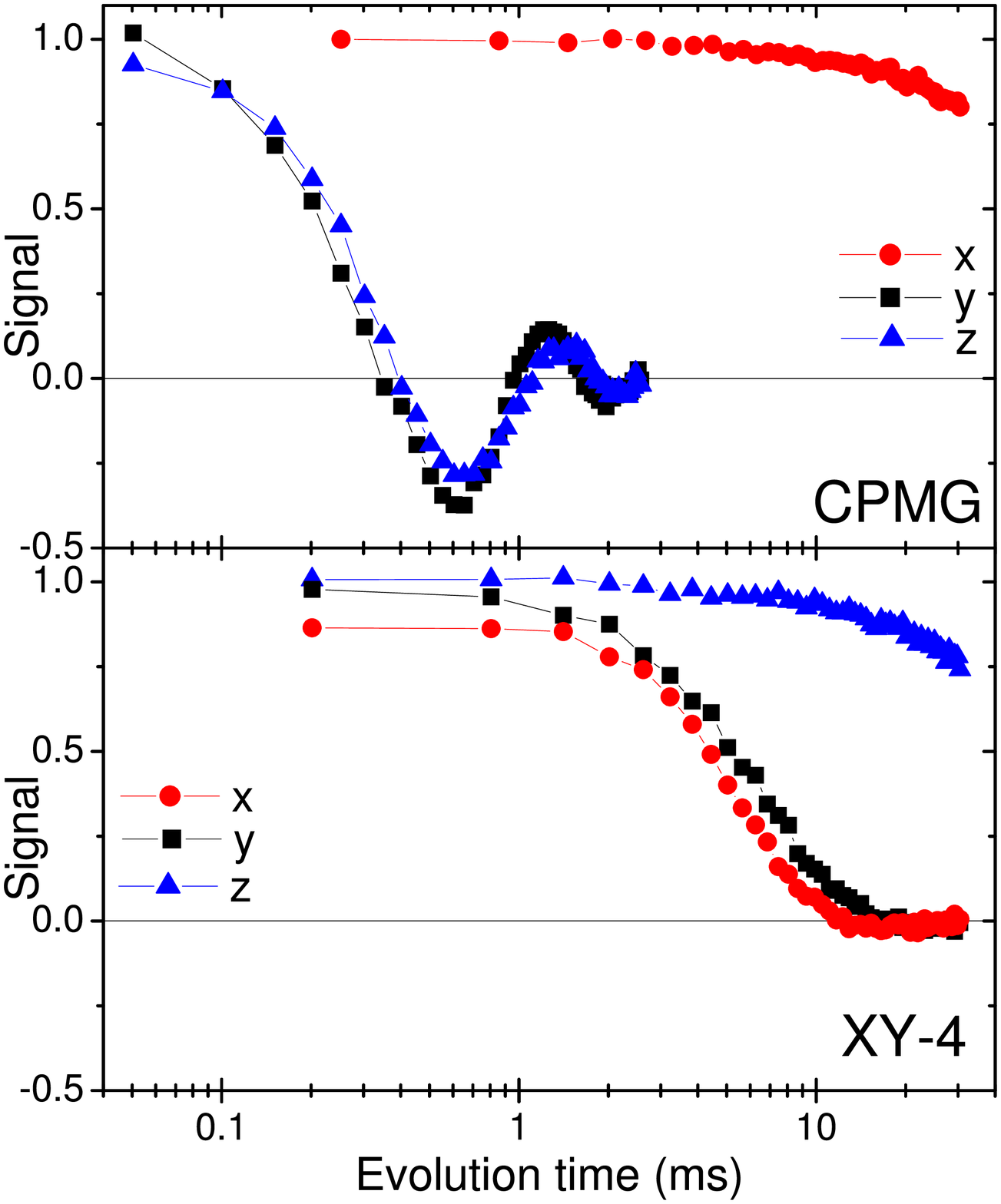}}} &
\subfigure[{\bf CPMG (1 cycle)}]{\includegraphics[width=4.5cm]{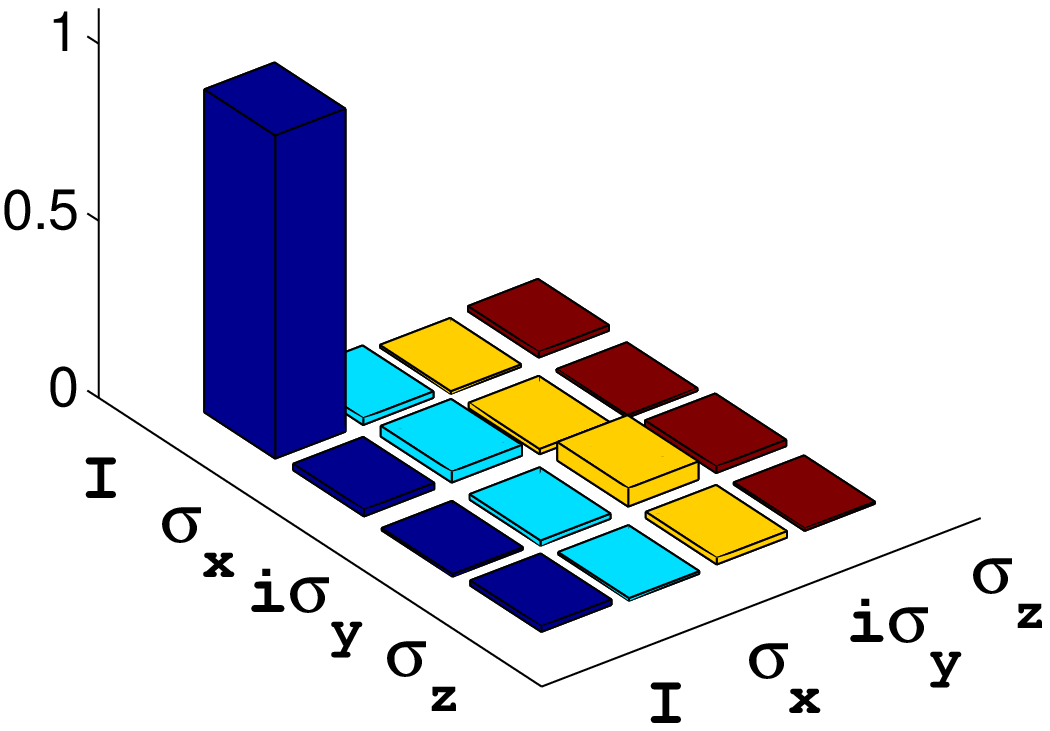}}&
\subfigure[{\bf CPMG (40 cycles)}]{\includegraphics[width=4.5cm]{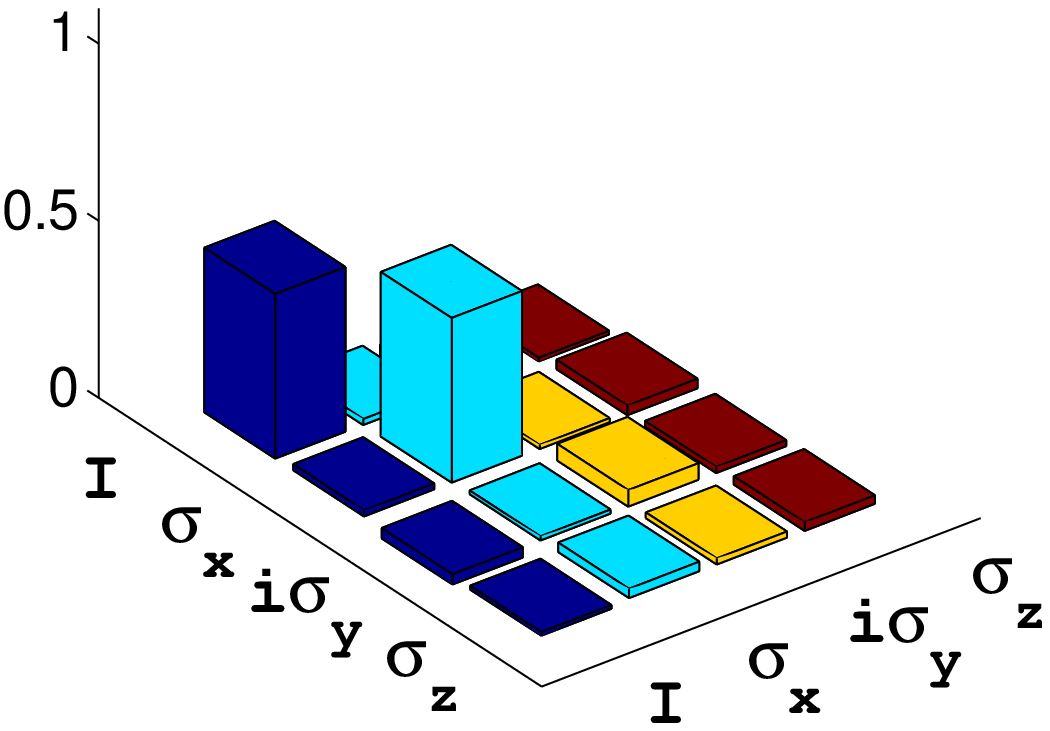}} \\
&\subfigure[ {\bf XY-4 (1 cycle)}]{\includegraphics[width=4.5cm]{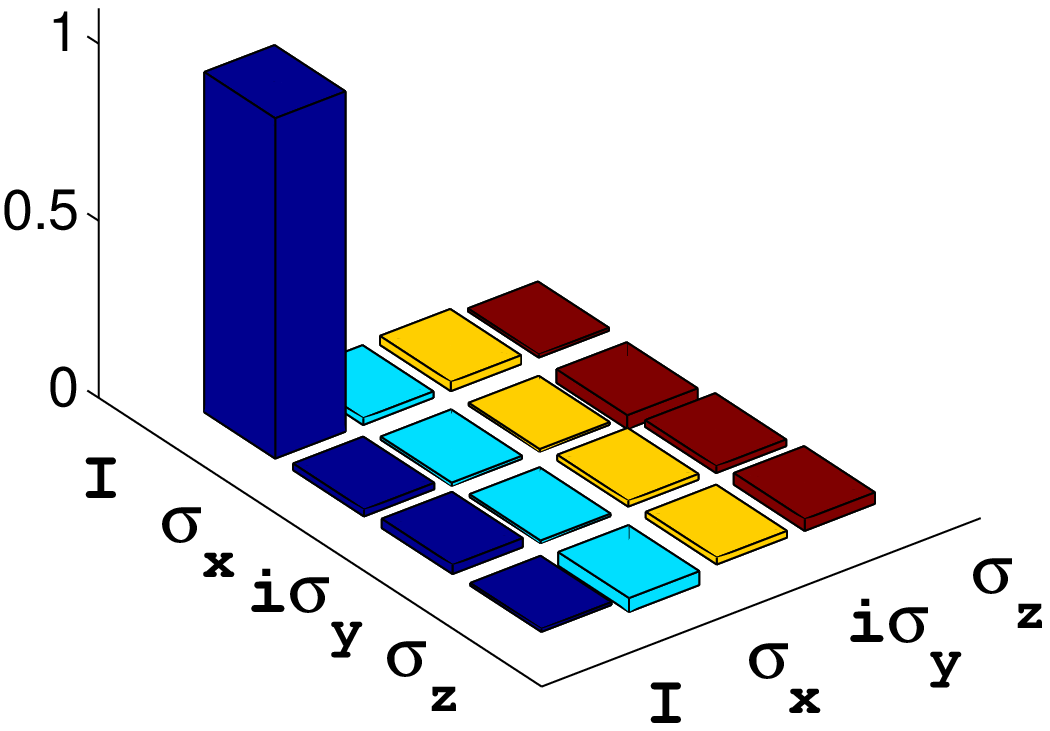}} &
\subfigure[ {\bf XY-4 (40 cycles)}]{\includegraphics[width=4.5cm]{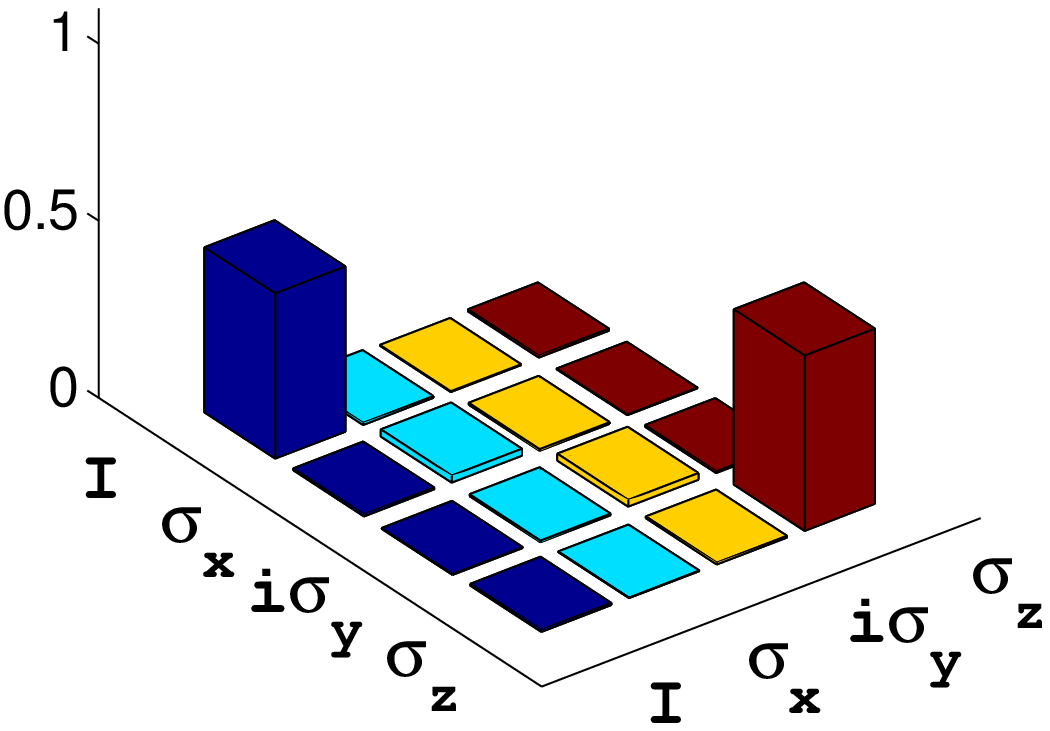}} 
\end{tabular}
\end{center}
\vspace*{13pt}
\caption{\label{chim} Comparison between two basic DD sequences: CPMG, which is  not robust
against errors,  and the self-correcting sequence XY-4. The left panel shows the normalized magnetization as a function of time.
The delay $\tau = 40 \mu \mathrm{s}$ between the pulses is constant and identical for both sequences. 
The panels on the left shows the real part of the process matrices $\chi$ for CPMG and XY-4. 
The imaginary part, which is very small, is not shown. }
\end{figure}

An example of the destructive effects of pulse imperfection 
is illustrated in the left panel of figure \ref{chim}.
Here, we measured  the 
magnetization decay of the $^{13}$C nuclear spins  during two different DD sequences. 
The sample used for this experiment was polycrystalline 
adamantane \cite{Alvarez2010,Ajoy2011,Souza2011,Souza2011a}. The 
dephasing of the nuclear spins originates from the
interaction with an environment consisting of $^1$H nuclear spins and can be considered 
as a pure dephasing process.  The first sequence 
considered in the figure is  CPMG. 
In this case we can observe that the decay of the magnetization is 
imperceptibly slow when the system is initially oriented parallel to the rotation
axis of the pulse (longitudinal state). 
As we discuss below, this is an indication that the pulse errors have no effect on this initial state.
In contrast, for a transverse initial state, the errors of the
individual pulses accumulate and lead to a rapid decay, as shown in Fig. \ref{chim}.
A similar behavior is found for the UDD sequence  \cite{Alvarez2010,Ajoy2011}.

The second DD sequence considered 
in  Fig. \ref{chim} is the XY-4 sequence, which consists of pulses applied
along the x and y axes. 
The alternating phases of the pulses results in a partial cancellation of 
pulse errors, independent of the initial condition \cite{Maudsley1986,Gullion1990}.
As a result, the performance of this sequence is much more symmetric with respect to 
the initial state in the xy-plane and the average decay times are significantly longer
\cite{Alvarez2010,Souza2011}.

In the context of quantum information processing, it is important that the performance of gate operations
be independent of the initial conditions (which typically are unknown). 
A common choice 
for quantifying the performance of a general quantum operation is then the fidelity $F$ \cite{Wang200858}:
\begin{equation}
F = \frac{|Tr(A B^{\dagger})|}{\sqrt{Tr(A A^{\dag}) Tr(B B^{\dag})}}.
\label{e:fid}
\end{equation}
Here, $A$ is the target propagator for the process and $B$ the actual propagator.
For the present situation, where the goal is a quantum memory, the target propagator
is the identity operation $I$.

We can not assume that the actual propagators are unitary.
We therefore write the process as
\begin{equation}
\rho_f = \sum_{nm} \chi_{mn} E_m \rho_i E_n^{\dag} 
\end{equation}
where $\rho_i$ and  $\rho_f$ are the density matrices at the beginning and end of the process.
The operators $E_m$ must form a basis. For the present case, we choose them as
$E_m = ( I,\sigma_x,i\sigma_y,\sigma_z )$. 
The ideal and actual processes can therefore be quantified by the matrix elements $\chi_{mn}$.
For the target evolution, the $\chi$-matrix is
\begin{equation}
\chi_I = \left( \begin{array}{cccc}
1 & 0 & 0 &  0 \\
0 & 0 & 0 &  0 \\
0 & 0 & 0 &  0 \\
0 & 0 & 0 &  0  
\end{array} \right).
\label{chiid}
\end{equation} 
 
The matrix elements for the actual process are determined experimentally 
by quantum process tomography \cite{qpt,Nielsen00}.
We use them to calculate the process fidelity from eq. (\ref{e:fid}).
In figure \ref{chim}, we compare the real part of the 
experimental $\chi$ matrices for the two sequences. 
After one cycle, the process matrices for both sequences 
represent a quantum operation that is close to the 
identity operation. 
The fidelity between the experimental matrices and the ideal 
matrix (\ref{chiid}) is $0.988$ and $0.989$ for the 
CPMG and XY-4 cycles, respectively. 
Measured over 40 cycles, the process matrices of the two sequences differ significantly 
from the identity operation but also from each order. 
For the XY-4 sequence, the non-vanishing elements are $\chi_{11}$ and $\chi_{44}$, while for 
CPMG the non-vanishing elements are  $\chi_{11}$ and $\chi_{22}$.  
The two matrices represent therefore 
qualitatively different processes. 
The XY-4 sequence 
destroys all transversal magnetization, in this case the 
resulting density matrix is
\begin{equation}
\rho_f = \chi_{11} \rho_i + \chi_{44} \sigma_z  \rho_i  \sigma_z.
\label{dmxy4}
\end{equation} 
 This corresponds to a projection of the density operator onto the $z$-axis,
 i.e. to a complete dephasing of the transverse components in the $xy$-plane.
 
The CPMG sequence, conversely, projects the density operator onto the $x$-axis:
\begin{equation}
\rho_f = \chi_{11}  \rho_i  + \chi_{22} \sigma_x  \rho_i  \sigma_x.
\label{dmcpmg}
\end{equation} 
The CPMG is a spin lock \cite{Santyr88} sequence that allows  retains magnetization in 
the $x$ direction but destroys all components perpendicular to it. 
Since a real experimental implementation always generates a distribution 
of control field amplitudes, spins at different positions precess with different rates
around the direction of the rf field.
As a result, the perpendicular components become completely randomized 
after a sufficiently large overall flip angle as shown in the left panel of Fig. \ref{chim}.

The performance of experimentally
accessible DD sequences is limited by the
pulse errors \cite{Alvarez2010,Ryan2010,Souza2011,Souza2011a}. In many situations, the dominating error contributions 
are flip angle and offset errors. In the next section we show different strategies to make 
decoupling sequences  robust against such errors.

\section{Robust decoupling sequences}

We have to make decoupling insensitive to pulse imperfections.
Different possibilities for generating high-fidelity sequences have been proposed
in the context  of quantum information 
processing \cite{khodjasteh_fault-tolerant_2005,khodjasteh_performance_2007,Souza2011}.
Here, we discuss two possible approaches: first we show that it is possible
to replace individual refocusing pulses by compensated pulses that implement
very precise inversions, and then we discuss sequences that are inherently robust,
i.e. insensitive to the imperfections of the individual pulses.

\subsection{Robust Pulses}

The simplest approach to make a sequence robust is by  
replacing  every standard pulse by a robust composite pulse \cite{composite}. 
The composite pulses are sequences of 
consecutive pulses designed to be robust against various classes of 
imperfections generating therefore rotations that are 
close to the ideal rotation even in the presence errors.  Particularly useful for 
quantum information applications are those composite pulses that 
produce compensated rotations for any initial condition, denominated in the 
NMR literature as class-A pulses \cite{composite}. 

Recent experiments have successfully used composite pulses to demonstrate
the resulting increase of the performance of different 
DD sequences \cite{Ryan2010,Souza2011}.  
These works have implemented the 
composite pulse defined as: 
\begin{equation}
(\pi)_{\pi/6+\phi}-(\pi)_{\phi}-(\pi)_{\pi/2+\phi}-(\pi)_{\phi}-(\pi)_{\pi/6+\phi},
\label{kpul}
\end{equation}
which is equivalent to a robust $\pi$ rotation around the axis defined by $\phi$ followed by a $-\pi/3$
rotation around the $z$ axis  \cite{kddtycko}. 
For cyclic sequences, which always consist of 
even numbers of $\pi$ rotations, the effect 
of the additional $z$ rotation vanishes  if the flip 
angle errors are sufficiently small.  

A comparison between standard sequences (not using robust pulses) against sequences with
robust pulses has been reported in \cite{Souza2011}. It was observed that robust pulses  
improve the performance at high duty cycles. However, for low duty cycles,
standard sequences perform better. This is due to the
shorter cycle time of the standard sequence if constant duty
cycles are compared. Thus, if the objective is only to preserve a quantum
state, the best performance is obtained at high duty cycles,
using robust pulses. The best sequences that are suitable for parallel application of quantum
gate operations are  the  self-correcting sequences discussed below.

The composite pulses are usually designed to correct flip angle errors and offset errors. For 
compensating the effects introduced by the finite length of pulses, some theoretical works have proposed
that a finite pulse could be approximated as an instantaneous one by using an appropriate  
shaped  pulse \cite{shape1,shape2,shape3}.  These works has provided analytical and numerical evidence 
that a careful shape  design  can strongly affects the performance 
of decoupling sequences. 

\subsection{Self-correcting sequences}

An alternative to the use of composite pulses consists in
making the decoupling sequences fault-tolerant without compensating the error of each pulse,
but by designing them in such a way that the error 
introduced by one pulse is compensated by the other 
pulses of the cycle \cite{Souza2011}. A straightforward strategy for 
designing improved sequences consists in concatenating one basic 
building block cycle into a longer and robust cycle.

The XY-4 cycle is often chosen as the  building block for constructing self-correcting 
sequences. 
This cycle is the shortest DD sequence that
cancels the zero-order average Hamiltonian for a general SE interaction 
and also has the advantage of being partially robust to pulse imperfections. 

\begin{figure}[htbp]
\vspace*{13pt}
\begin{center}
\includegraphics[width=10.5cm]{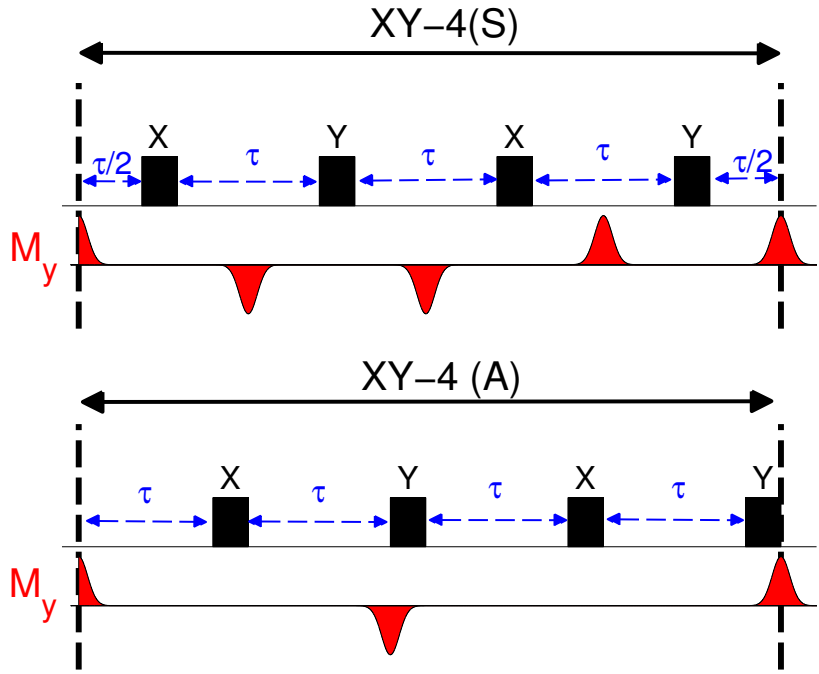}
\end{center}
\vspace*{13pt}
\caption{\label{ecos}  Schematic representation of time
symmetric XY-4(S) and asymmetric XY-4(A). } 
\end{figure}

In the spectroscopy and 
quantum computing communities, two versions of the XY-4 sequence are used \cite{Souza2011a}. 
The basic cycle 
originally introduced in NMR  shows reflection symmetry with respect to the 
center of the cycle. In contrast 
to that, the sequence used in the quantum information community is 
time asymmetric. One consequence of this small difference is 
that in the symmetric version, the echoes are formed in the center of 
the windows between any two pulses,
while in the case of asymmetric cycles, the echoes coincide with every 
second pulse,  as shown in figure \ref{ecos}. 
The separation in time between the echoes is therefore twice as long in 
this case. If the environment is not static, the larger separation of the echoes leads to 
a faster decay of the echo amplitude \cite{Souza2011a}.

If a robust pulse only contains $\pi$ rotations, as in the case of (\ref{kpul}), it 
is also possible to convert such a composite pulse into a decoupling cycle
by inserting delays between the individual $\pi$ rotations.
This approach has been used in \cite{Souza2011} to build 
an self-correcting sequence. 
The basic cycle is then
\begin{equation}
 KDD_\phi = f_{\tau/2}(\pi)_{\pi/6+\phi}f_{\tau}(\pi)_{\phi}f_{\tau}(\pi)_{\pi/2+\phi}f_{\tau}(\pi)_{\phi}f_{\tau}(\pi)_{\pi/6+\phi}f_{\tau/2}. 
\end{equation}
The self-correcting sequence is created by combining 5-pulse 
blocks shifted in phase by $\pi/2$,
such as [KDD$_\phi$ - KDD$_{\phi+\pi/2}$]$^2$, where the lower index
gives the overall phase of the block. The cyclic 
repetition of these 20 pulses is referred to as the KDD sequence \cite{Souza2011}. 

\begin{figure}[htbp]
\vspace*{13pt}
\begin{center}
{\includegraphics[width=12.5cm]{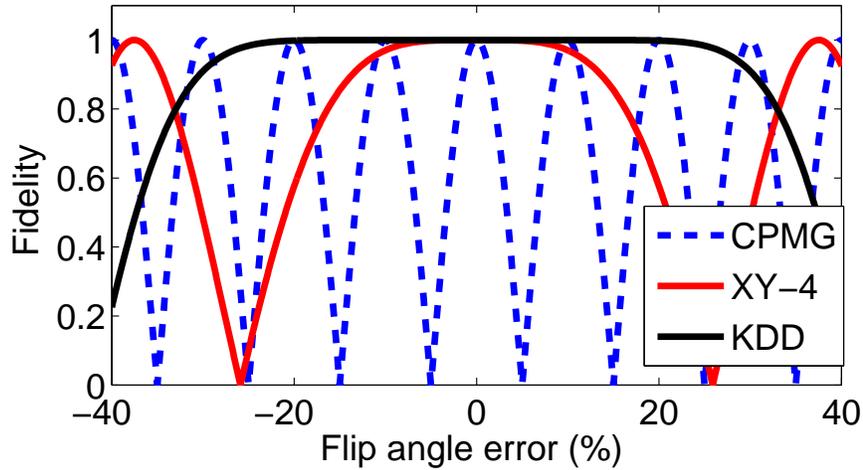}}
\end{center}
\vspace*{13pt}
\caption{\label{fidelity} Simulation of fidelity as a function of the 
flip angle error for KDD, XY-4 and CPMG cycles. }
\end{figure}

In Fig. \ref{fidelity}, we show how strongly  errors in the flip angles of individual pulses
affect the fidelity of the pulse sequence.
Neglecting the effect of the environment, we calculate the 
fidelity $F$ after the application of 20 pulses as a function of the flip 
angle error. 
The figure compares the fidelities  for the CPMG, XY-4 and KDD cycles. 
For the CPMG sequence, the fidelity drops to <95\% if the flip angle error
exceeds $\approx$ 2 \%.
For the XY-4 sequence, this bandwidth increases to $\approx$ 10 \% and for 
KDD to $\approx$ 30 \%.
KDD and XY-4 are obviously much less susceptible to pulse imperfections than CPMG. 
The low fidelities observed for CPMG is experimentally 
manifested by the fast decay of the transverse components, such as $M_x$ in Fig. \ref{chim}. 

\subsection{Combining basic cycles}

Every decoupling sequence contains unwanted terms in the average Hamiltonian.
They can be reduced by combining different versions of the basic cycles in such a way
that some of the error terms cancel.
Two different versions of this procedure have been used:
The basic cycles can be applied subsequently \cite{Gullion1990} or one cycle can be inserted
into the delays of another cycle \cite{khodjasteh_fault-tolerant_2005,khodjasteh_performance_2007}.
The first approach was introduced in NMR, e.g. for designing high-performance
homonuclear decoupling sequences  \cite{waugh68b,mansfield73,rhim73,rhim74,burum79,Burum81}
 or in high-resolution 
heteronuclear decoupling \cite{Hahn50b,Carr1954,Meiboom1958,Maudsley1986,Gullion1990,waugh69}.
Examples of DD sequences that can be constructed using 
this approach are XY-8 and XY-16 sequences \cite{Gullion1990}. 
Here the XY-8 is created 
combining a XY-4 cycle with its time-reversed image while XY-16 is created combining 
the XY-8 with its phase-shifted copy.

The second approach is the concatenation scheme 
proposed by Khodjasteh and 
Lidar \cite{khodjasteh_fault-tolerant_2005,khodjasteh_performance_2007}. 
It generates the  CDD sequence of order $n+1$ by inserting  CDD$_n$ cycles into the 
delays of the XY-4 sequence (see figure \ref{ddseqs}). 
Ideally, each level of concatenation improves the decoupling performance and the tolerance to 
pulse imperfections; in practice,  higher order sequences
do not always perform better.  
It has been  theoretically predicted 
\cite{Viola2003,khodjasteh_performance_2007,Hodgson2010} and 
later observed experimentally that 
the finite duration of the pulses and constrained delays between pulses result in the existence of
optimal levels of concatenation \cite{Alvarez2010,Souza2011}.  

\begin{figure}[htbp]
\vspace*{13pt}
\begin{center}
{\includegraphics[width=14.5cm]{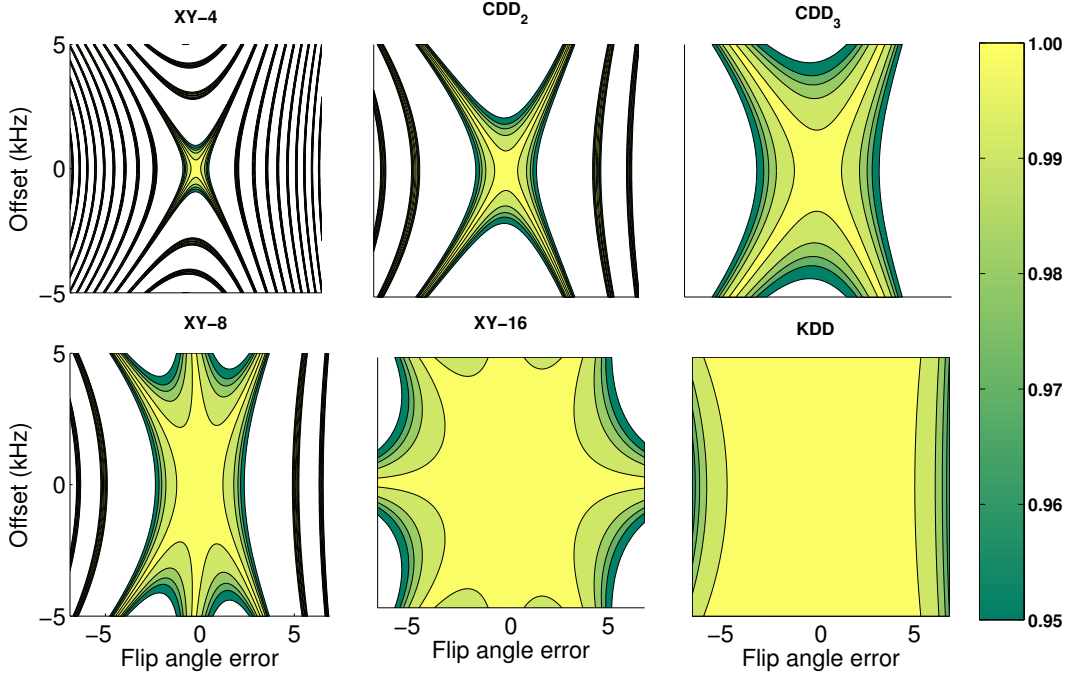}}
\end{center}
\vspace*{13pt}
\caption{\label{sfseqs} Error tolerance of different self-correcting sequences.
The upper row shows the calculated fidelity F for CDD sequences, while the lower 
row shows the results for XY-8, XY-16 and KDD sequences.  Each panel shows
the fidelity after 1680 pulses as a function of flip-angle error and
offset errors. The regions where the fidelity is lower than 0.95 are
shown in white.}
\end{figure}

In figure \ref{sfseqs}, we demonstrate how the well-designed combination of basic cycles
can lead to extended cycles with better error compensation.
Here, we consider as the leading experimental imperfections deviations of the amplitude 
and frequency of the pulse. 
Neglecting the effect of the environment, we calculate the fidelity $F$ after applying 1680 
pulses to the system as a function of the two error parameters.
Each panel contains the color-coded fidelity for one of six different decoupling sequences.
In the top panels we can clearly see the improvement in the error tolerance due 
to the CDD scheme of concatenation.  
Panels 1, 4 and 5 show the same result for the sequential concatenation scheme, 
where only two cycles are combined at each step:
concatenation of the XY-4 cycle with its 
time-inverted and phase-shifted copies  forms the XY-8 
and XY-16 sequences. 
The 16-pulse XY-16 cycle is significantly more robust than the 84-pulse CDD$_3$ cycle.
The best performance is achieved by the KDD sequence, whose cycle consists of 20 pulses .

\begin{figure}[htbp]
\vspace*{13pt}
\begin{center}
{\includegraphics[width=10.5cm]{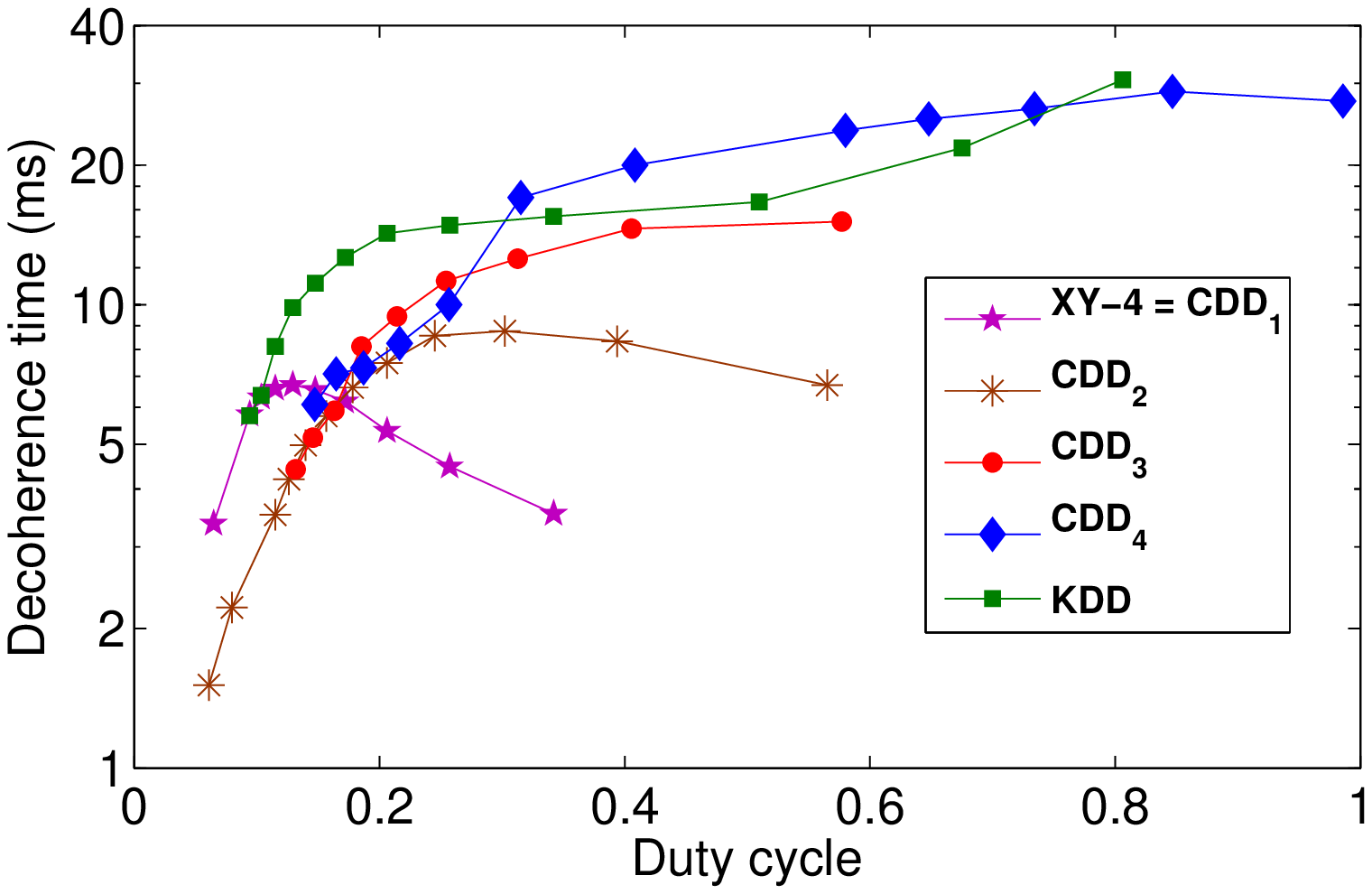}}
\end{center}
\vspace*{13pt}
\caption{\label{kddfig} Experimental decoherence times for different 
compensated DD sequences as a function of the duty cycle.}
\end{figure}

In Fig. \ref{kddfig}, we compare the experimental performance of different self-correcting sequences.
The performance of the CDD sequences 
always saturates or decreases with increasing duty cycle under 
the current experimental conditions \cite{Souza2011}. However, instead of saturating, the relaxation
time for the KDD sequence continues to increase, as in the case of sequences with
robust pulses \cite{Souza2011a}. 
The KDD sequence combines the useful properties of robust
sequences with those of sequences of robust pulses and can thus be used
for both quantum computing and state preservation.

\subsection{Time reversal symmetry \label{symmetry}}

\begin{figure}[htbp]
\vspace*{13pt}
\begin{center}
\subfigure{\includegraphics[width=5.5cm]{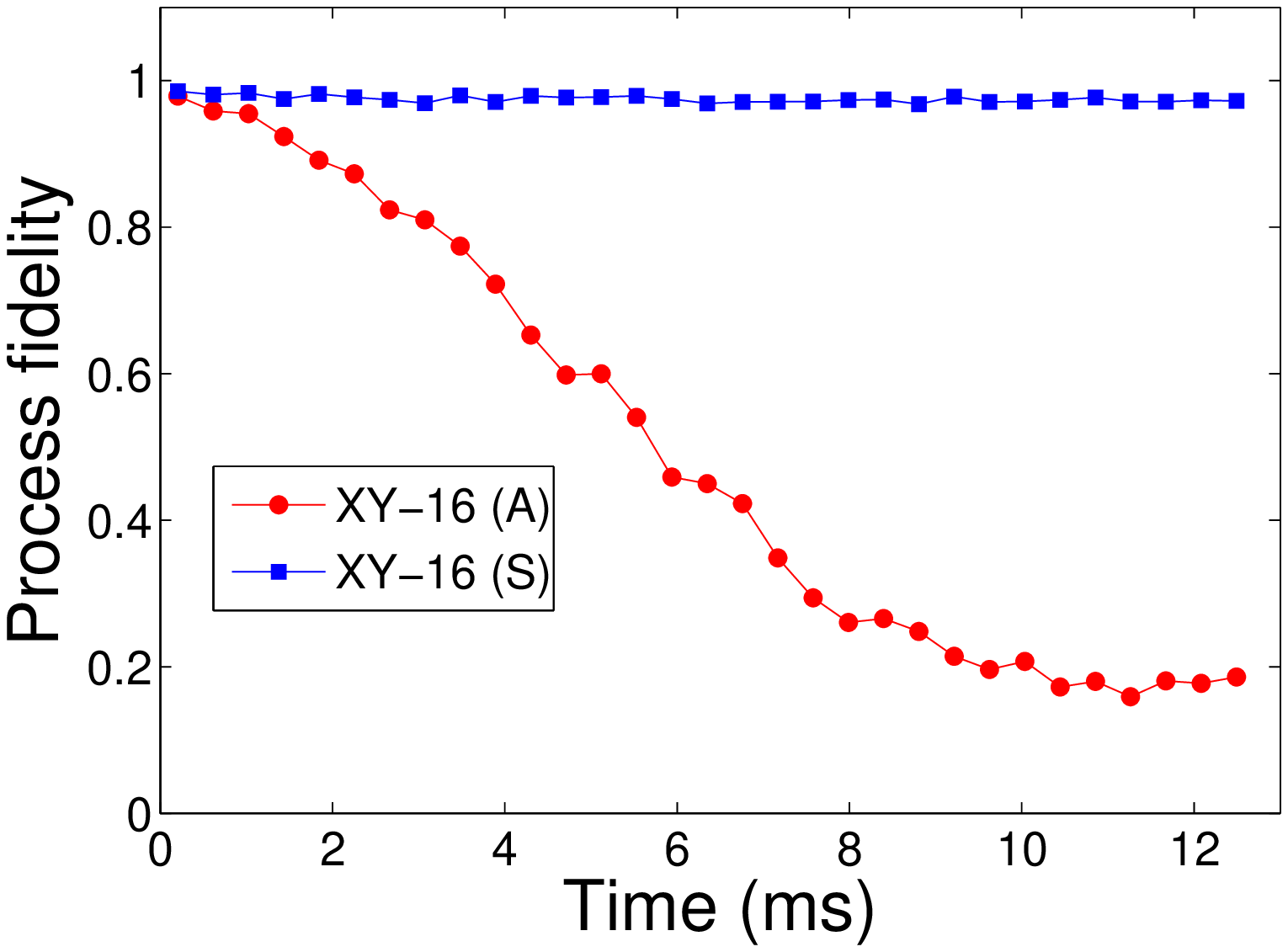}}
\subfigure{\includegraphics[width=8.0cm]{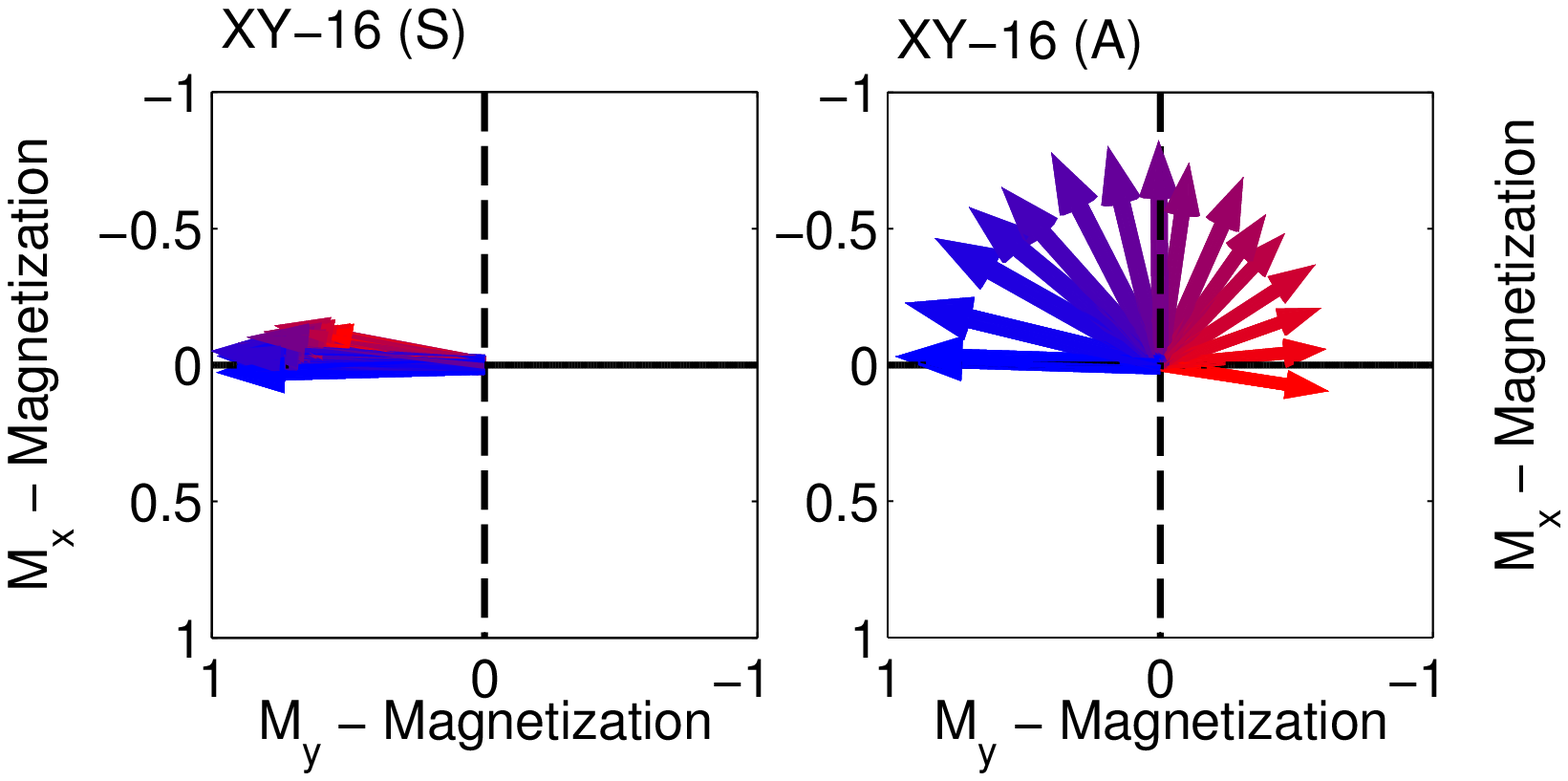}}
\end{center}
\vspace*{13pt}
\caption{\label{fidelxy}  Comparison between the two forms of XY-16 sequence: XY-16 (S), built 
from symmetric form of the XY-4 cycle, and XY-16 (A), built from the asymmetric block. The 
delay $\tau = 10 \mu \mathrm{s}$ between the pulses is constant and identical for both sequences. The 
left panel shows the process 
fidelity as a function time. The right panel shows the Bloch 
vector in the xy plane at different times. The color code in the right panel denotes 
the time evolution, blue for the initial state and red for the final state. } 
\end{figure}

The symmetry of the basic building blocks has a key role in determining the performance
of the concatenated higher order sequences. 
Two sequences constructed
according to the same rules from a basic block have different propagators if
the basic blocks are symmetric or not \cite{Souza2011,Souza2011a}. 
If we concatenate four 
XY-4 cycles to the XY-16 sequence, e.g., we obtain new cycles,
which are  time-symmetric, independent of which
version of the XY-4 sequence was used for the building blocks. 
Although all the odd order terms 
vanish in the average Hamiltonians of both versions, the even order terms of 
the sequences that are built from asymmetric blocks contain additional 
unwanted terms \cite{Souza2011a}. The different behavior of sequences consisting of symmetric 
vs. asymmetric blocks is illustrated in figure \ref{fidelxy}. 
If we start from the symmetric form 
of XY-4, the resulting XY-16 sequence shows much better performance than the sequence using the asymmetric 
XY-4 as the building block. 
Analogous results were obtained for the two versions of the XY-8 sequence \cite{Souza2011a}.

Earlier experiments showed two different 
contributions to the overall fidelity loss \cite{Souza2011a}: 
A precession around  the $z$-axis, which can be attributed 
to the combined effect of flip-angle errors
and an overall reduction of the amplitude, which results from the 
system-environment interaction. The combination of precession and reduction of amplitude 
is illustrated in right panel of Fig. \ref{fidelxy}. In this figure we show the xy-components 
of the magnetization at different times during the XY-16 sequence. If the XY-16 sequence 
is  built by the asymmetric form of XY-4 a distinct precession around 
the $z$-axis  is observed. 
This causes a deviation from the desired evolution and reduces therefore the 
fidelity of the process. 
However, for the sequence consisting of symmetric 
blocks, the precession is negligible. 
These results suggest that  pulse errors are  better
compensated by concatenating symmetric building blocks.

\begin{figure}[htbp]
\vspace*{13pt}
\begin{center}
\includegraphics[width=12.5cm]{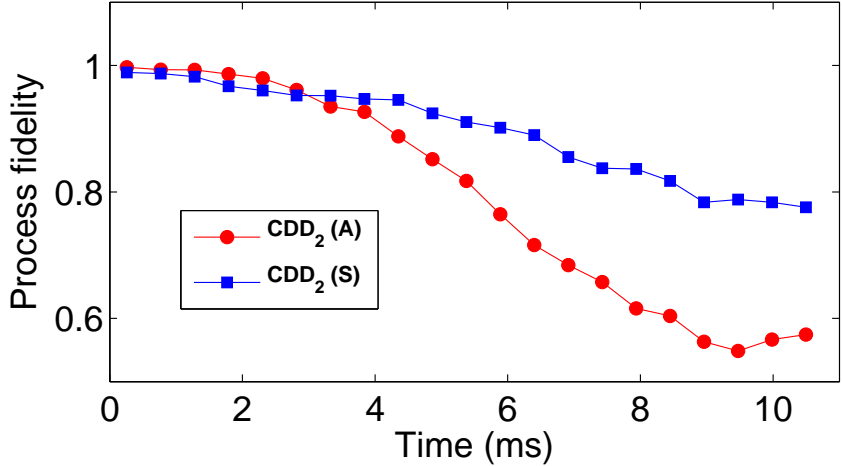}
\end{center}
\vspace*{13pt}
\caption{\label{fcdd}  Experimental fidelity decay for CDD$_2$ built from 
symmetric and asymmetric blocks. The delay $\tau = 10 \mu \mathrm{s}$ between the pulses 
is constant and identical for both sequences.  } 
\end{figure}

The same concept can also be applied to CDD sequences. The
conventional concatenation scheme of eq. (\ref{cdd}) uses asymmetric building blocks and is not 
compatible with the symmetric version of XY-4. 
A new concatenation 
scheme was therefore proposed in \cite{Souza2011,Souza2011a}. 
In this scheme, the  symmetrized 
version of CDD is constructed  as 
\begin{eqnarray}
CDD_{n+1} = [\sqrt{CDD_{n}}-X-CDD_{n}-Y-\sqrt{CDD_{n}}]^2 .
\end{eqnarray}
In Fig. \ref{fcdd} we compare the process fidelities for the 
two versions of the CDD$_2$ sequence. As in the case of XY sequences, clearly, the symmetrized 
version, CDD$_2$(S), shows a significantly 
improved performance, compared to the standard CDD$_2$(A) version. In Ref. \cite{Souza2011a}, it was 
experimentally observed that the performance of all DD sequences based on symmetric building blocks is 
 better or equal to that of sequences using non-symmetric 
building blocks.
This behavior is consistent with general arguments based on average 
Hamiltonian theory \cite{levitt1,levitt2}.

\section{Conclusion and perspectives} 

Dynamical decoupling is becoming a standard technique for preserving the coherence of 
quantum mechanical systems, which does not need control over the environmental degrees of freedom. 
The technique aims to
reduce decoherence rates by attenuating the system-environment
interaction with a periodic sequence of $\pi$ pulses applied to the qubits. The pioneering 
strategies 
for decoupling were introduced in the context of NMR spectroscopy \cite{Hahn50b}. 
Since then,  many 
different decoupling sequences have been developed
in the context 
of NMR  \cite{Carr1954,Meiboom1958,Maudsley1986,Gullion1990} or quantum information 
processing  
\cite{viola_dynamical_1999,khodjasteh_fault-tolerant_2005,khodjasteh_performance_2007,Uhrig2007,Uhrig2009,west_near-optimal_2010,qdd1,qdd2,Souza2011}.

Generally, we can divide the DD sequences in two groups: (i) Sequences
that involve pulses in a single spatial direction and (ii) sequences
that contain pulses in different spatial directions. 
The type (i) sequences
are strongly sensitive to pulse errors and are only capable of suppressing the effects of a purely dephasing
environment or pure spin-flip interaction. 
Examples of 
such DD sequences are CPMG and UDD. The second group can suppress a general SE interaction and usually 
exhibits better tolerance to experimental imperfections. 
Example of such sequences are the 
XY-family (XY-4, XY-8 and XY-16), the  CDD sequences and the KDD sequence. 

Recent experiments have
successfully implemented DD methods and demonstrated
the resulting increase of the coherence times by several orders of magnitude \cite{Alvarez2010,Ryan2010,Souza2011,Souza2011a}. 
These works also
showed that the main limitation to the reduction of the decay rates are the imperfections of the  pulse. 
Two approaches have been used to correct this.
The first approach replaces the inversion pulses by robust composite pulses \cite{composite},
which generate rotations that are close to the target value even in the presence of pulse errors. 
In this case, the
pulses are corrected individually. 
The second approach consists in
designing the decoupling sequences in such a way that the
error introduced by one pulse is compensated by the other
pulses, without compensating the error of each pulse individually. 
The properties of basic decoupling cycles can be further improved by concatenating  basic 
cycles into longer and more robust cycles. 
The concatenation can be made either by
combining symmetry-related copies of a basic cycle subsequently \cite{Gullion1990} (resulting in the 
XY-8 and XY-16 sequences) or by inserting
the basic cycle into the delays of 
another cycle \cite{khodjasteh_fault-tolerant_2005,khodjasteh_performance_2007} (CDD sequences).

The time reversal symmetry of the basic building blocks is a useful criterion for minimizing
error contributions. It has been demonstrated that the sequences
built from symmetric building blocks often perform better and never worse than 
sequences built from non-symmetric blocks \cite{Souza2011,Souza2011a}. 
This is a significant advantage,
considering that the complexity of the sequences based on symmetric or asymmetric blocks 
are identical.

Earlier experiments \cite{Souza2011} showed  that the best sequences that are suitable for 
parallel application of quantum
gate operations are the symmetric self-correcting sequences.
However, as the delay between pulses decreases, sequences with robust pulses 
perform better.
Thus, if the objective is only to preserve a quantum
state, the best performance is achieved by using robust pulses to correct pulse errors. On the 
other hand, the KDD sequence introduced in  \cite{Souza2011}  combines the useful properties 
of self-correcting sequence with those of robust pulses and can thus be used
for both quantum memory and quantum computing.

During the last years, many advances have been achieved. However, for the application of the 
technique in real quantum devices, further 
studies will certainly be required. 
So far, most work has focused on  single qubit 
systems. 
In the future, more experimental tests will be needed with  
multi-qubit systems. In the field of quantum computation, another important 
development  may result from 
the combination of dynamical decoupling sequences with those techniques used to implement 
robust quantum gates \cite{error1,6581,error3,error4}.  Since DD does not require
auxiliary qubits or measurements, it can be used as an
economical alternative to complement quantum error correction. 
Some theoretical works \cite{ddgate1,ddgate2,ddgate3,ddgate4} proposed methods for 
combining the two methods but no experimental 
test was carried out to date. Future 
research on dynamical decoupling will also focus on applications outside of quantum information processing. 
Recent experiments have  applied DD pulse sequences, for example, to probe the 
noise spectrum directly \cite{noise1,noise2,Alvarez2011a} and 
detect  weak magnetic fields  \cite{magnetometer1,meriles,magnetometer2,magnetometer3}.

\begin{acknowledgments}
We acknowledge useful discussions with Daniel Lidar and Gregory Quiroz. This work is supported by the
DFG through Su 192/24-1.
\end{acknowledgments}


\providecommand{\noopsort}[1]{}\providecommand{\singleletter}[1]{#1}%

\end{document}